\documentclass[pra,twocolumn,showpacs,amsmath]{revtex4}

\usepackage{graphicx}
\begin{document}

\title{Level crossings in a cavity QED model}

\author{J.~Larson}

\affiliation{Physics Department\\
       Royal Institute of Technology (KTH)\\
       Albanova,
       Roslagstullsbacken 21\\
       SE-10691 Stockholm,
       Sweden}

\date{\today}

\begin{abstract}
In this paper I study the dynamics of a two-level atom interacting with a standing wave field. When the atom is subjected to a weak linear force, the problem can be turned into a time dependent one, and the evolution is understood from the band structure of the spectrum. The presence of level crossings in the spectrum gives rise to Bloch oscillations of the atomic motion. Here I investigate the effects of the atom-field detuning parameter. A variety of different level crossings are obtained by changing the magnitude of the detuning, and the behaviour of the atomic motion is strongly affected due to this. I also consider the situation in which the detuning is oscillating in time and its impact on the atomic motion. Wave packet simulations of the full problem are treated numerically and the results are compared with analytical solutions given by the standard Landau-Zener and the three-level Landau-Zener models. 
\end{abstract}


\maketitle

\section{Introduction}\label{intro}
The general theory of systems evolving with a periodic Hamiltonians has long been known and understood. However, it is only recently that some of the theoretical predictions  have been tested experimentally. It has turned out that the physical systems one first  had in mind, namely these of solid state materials, have several drawbacks making it hard to realize such experiments. Instead of using the periodic structure in crystal lattices, one may use a standing wave light field as an 'effective' potential, acting upon the internal electron states of an atom. Here the wavelength of the field sets the period, and due to lack of lattice defects in the standing wave and long atomic decay times, the decoherence times exceed those in solid state systems by several orders of magnitude. One purely quantum mechanical effect of the atomic motion is seen when the atom is subjected to a constant force $F$. The optical potential then has a washboard shape. As long as the force is weak compared to the amplitude of the standing wave, the quantum numbers that describe the system are still good numbers, apart from the fact that the quasi momentum will follow the classical evolution $k=k_0-Ft$. Thus the system remains in its initial energy band while it sweeps the quasi momentum. Since the energy bands are periodic in $k$, the motion will also be periodic contrary to the constantly accelerating one expected from classical mechanics. This phenomena is called Bloch oscillations after the investigations \cite{bloch1a,bloch1b}. These oscillations have been verified experimentally in various systems; atoms in optical potentials \cite{atombloch1,atombloch2,atombloch3}, BEC's in optical lattices \cite{bec1,bec2} and solid state superlatticies \cite{super}. Some theoretical studies are presented in refs.~\cite{bloch2,lzbreak}, which analyze the problem from a Floquet or Bloch theory point of view. The dynamics can also be understood using a Wannier-Stark ladder spectrum, consisting of equally spaced discrete complex eigenvalues \cite{wannier}. 

With no atom-field coupling, the atom moves freely and the spectrum, consisting of an infinite set of bare energy dispersion curves, is continuous $E>0$. When the periodic potential is turned on, the degeneracy points where the curves cross are split and gaps are formed. The dispersion curves become periodic in $k$, with a period equal the Brillouin zone. The avoided crossings are central for the evolution when $k$ grows linearly in $t$. By linearelizing the dispersion curves around a crossing, the system serves as a testing ground for the Landau-Zener model \cite{lz1,lz2}. This model treats a linear time-dependent two-level crossing system analytically, and the asymptotic solutions for transitions between the two states are given. The Landau-Zener tunneling has been tested experimentally in connection with Bloch oscillations \cite{atombloch2,bec2,lztun} and also been the subject of theoretical investigations in the same system \cite{lzbreak}. 

In all the works mentioned above, the particle lacks internal structure. For example, in the situation of atoms interacting with a standing wave field, it is assumed that the detuning between the atomic transition frequency involved and the field frequency is large, implying that one of the two atomic levels can be omitted after an adiabatic elimination. This results in a two photon process and no single photon exchanges are present. The main reason for using a large detuning is that only the ground state and not the upper atomic level is populated and the coherence times increase considerably. In the experiments, a strong driven light field is used making the decay of the field negligible.     
  
In this paper I investigate the situation when the detuning may be small, which is fairly unanalyzed, one reference is \cite{zerodet}. It is understood that the paper is mostly of  theoretical interest, neglecting decays of both the atom and the field. I do, however, mention how to overcome the problem of losses. The field may be driven by an external classical source, and one may use a $\Lambda$-type of atom where one of the lower states is coupled to the upper state by the standing wave and the other lower state is coupled to the upper one by an external classical traveling wave field. If the upper state can be adiabatically eliminated, the effective two-level system has the same form as the one used in this paper, and thus the losses due to atomic excitations can be minimized. I argue how an effective model for describing losses of the field may be derived in certain case, and numerically show the effect of the losses. I keep the field quantized, assuming the field to be in an $n$ photon Fock state; this does not make the analysis more complicated than for a classical field. In contrast to the paper \cite{zerodet} I consider not only the zero detuning situation. By controlling the detuning, a new type of level crossing is achieved where three bare energy curves cross. This kind of crossing have similarities with the three-level Landau-Zener model \cite{lz3}, and comparisons between wave packet simulations of the full atom-field system and the analytically solvable three-level Landau-Zener model is carried out. Thus, this atom-field model provides one physically interesting system where this three-level Landau-Zener generalization is applicable. 

I also investigate the situation when the detuning becomes time-dependent and particularly when it oscillates in time, which is the situation when the atom is strongly Stark shifted by an alternating external electric field. In these considerations, the force is not included, and hence the dependence of time arise from the oscillating detuning only. The atom again traverses a level crossing and is, in the adiabatic limit, transferred between different bare states, and, consequently,  its motion may have an oscillating behaviour. 

The paper is organized as follows. In section \ref{sec2} the extended Jayness-Cumings Hamiltonian describing the system without an external constant force is introduced. I discuss general features like the two natural basis states, the band structure and effective parameters. The zero detuning case with the force applied is considered in \ref{sec3}, and I show the results from numerical wave packet simulations. The Landau-Zener transitions and the Bloch oscillations are discussed. The three-level crossing model is presented in section \ref{sec4}, and the wave packet simulations are compared with the analytic results of paper \cite{lz3}. In the next section \ref{sec5}, I look at the situation when there is no force acting on the system, but the detuning is oscillating in time. It is found that a similar evolution may be obtained as in the case with a force. A discussion about how to minimize losses, of both the atom and the field subsystem is considered in section \ref{sec6}, where I also show the effect of field losses on the atomic inversion. Finally I conclude the paper with conclusions in \ref{concl}.

\section{The model}\label{sec2}
The evolution of the combined atom plus a cavity mode is described by a Jaynes-Cummings type of interaction \cite{jc1,jc2}, where the coupling is $x$-dependent following from the variation of the electric field $\bar{E}(x)$ along the cavity mode, $g(x)=\bar{d}\cdot\bar{E}(x)/\hbar$. In particular, the mode is assumed to have a standing wave shape, $g(x)=2g_0\cos(qx)$. For cold atoms, their kinetic energy term must be taken into account fully quantum mechanically \cite{mazer0,refl1}, and we arrive at
\begin{equation}\label{ham}
H=-\frac{\hbar^2}{2\tilde{m}}\frac{d^2}{d\tilde{x}^2}+\!\frac{\tilde{\Delta}}{2}\sigma_z+\tilde{g}_02\cos(q\tilde{x})\left(a^{\dagger}\sigma^-+a\sigma^+\right)\!,
\end{equation}
where $m$ is the atomic mass, $\tilde{\Delta}$ is the atom-field mode detuning, $a^{\dagger}$ ($a$) are raising (lowering) boson operators for the mode state, $q=2\pi/\lambda$ is the photon wave number and $\sigma_z$, $\sigma^\pm$ are the Pauli $z$, raising/lowering operators acting on the two-level atom. The dynamics governed by (\ref{ham}) has been studied in numerous papers, see {\it i.e} \cite{dyn1,dyn2,dyn3,dyn4,dyn5,dyn6,dyn0}.  

In the following we will use scaled units. The photon recoil energy $E_R=\hbar^2q^2/2\tilde{m}$ defines a characteristic energy scale and $q^{-1}$ sets a length scale. The new scaled parameters are
\begin{equation}
\begin{array}{lllll}
x=q\tilde{x} & & p=\frac{\tilde{p}}{q} & & t=\tilde{t}\frac{E_R}{\hbar} \\ \\ E=\frac{\tilde{E}}{E_R} & &V_0=\frac{2\tilde{g}_0}{E_R} & & \delta=\frac{\tilde{\Delta}}{E_R},
\end{array}
\end{equation}
where the tilde $\sim$ indicates unscaled quantities.   

The scaled Hamiltonian has a spatial period $d=2\pi$, and according to Bloch or Floquet theory, it is known that the system eigenstates are determined by two quantum numbers; a discrete number $\nu$ and a continuous number $k$ \cite{dyn3}. The first $\nu$ is the band index ranging from $1,2,...$ and $k$ is the quasi momentum taking values within the first Brillouin zone $-1<k<1$ (scaled units). In some papers, the Brillouin zone has been defined as $-1/2<k<1/2$, with the consequence that two sets of energy bands are obtained, one for each internal state $|\pm\rangle$ \cite{dyn1,zerodet}. The spectrum $E_{\nu}(k)$ forms an infinite set of continuous bands separated by gaps. For smooth potentials, the gap size decreases for increasing values of $\nu$; in general the band-to-gap ratio falls off faster than $1/\nu$. Since upper $|\uparrow\rangle$ and lower $|\downarrow\rangle$ electronic states of the atom couple through absorption or emission of one photon, the number of excitations of the system is preserved by the Jaynes-Cummings type of interaction (\ref{ham}). We may thus, for a given number of excitation, define the excitation eigenstates
\begin{equation}
\begin{array}{c}
|+\rangle=|n-1\rangle|\uparrow\rangle \\ 
|-\rangle=|n\rangle|\downarrow\rangle,
\end{array}
\end{equation}
where $|n\rangle$ is the $n$ photon number state of the mode. These will also be referred to as internal states. In each exchange of energy between the field and the atom, the atomic momentum is shifted by $\pm1$ (in scaled units) from the mechanical 'kick' due to absorption or emission of a photon. Consequently, an internal state, $|+\rangle$ ($|-\rangle$), will only return to the same internal state after having shifted the momentum by an even number of momentum kicks. A state with momentum $p=k$, $|k\rangle$, and internal state $|-\rangle$, is coupled to the following set of states
\begin{equation}\label{bare}
|\psi_\mu(k)\rangle=\left\{
\begin{array}{ll}
|k+\mu\rangle|-\rangle, & \mu\,\,\,\mathrm{even} \\
|k+\mu\rangle|+\rangle, & \mu\,\,\,\mathrm{odd},
\end{array} \right.
\end{equation}
where $\mu$ runs over all integers. Note that the momentum of $|\psi_\mu(k)\rangle$ is $k+\mu$, and the quasi momentum $k$ will therefore be imagined within the first Brillouin zone. There is an orthogonal set of states with even/odd reversed, which is obtained by using the initial state $|k\rangle|+\rangle$. These two sets do not couple, and in the special case of zero detuning $\delta=0$ their spectra become identical. This degeneracy will, however, never be interesting for us, since we will always assume our system to be in one of the two sets. The states of Eq.~(\ref{bare}) are egenstates of the Hamiltonian in the absence of coupling $g_0=0$, and I will refer to them as {\it bare states}. For a non-zero coupling the bare states are no longer eigenstates of the Hamiltonian, and the new eigenstates, 
\begin{equation}
\begin{array}{lll}
H|\phi_\nu(k)\rangle=E_\nu(k)|\phi_\nu(k)\rangle, & & \nu=1,2,3,...\, ,
\end{array}
\end{equation}
will be called {\it dressed states}. In the theory of periodic operators, the dressed states are also known as Bloch or Floquet states and their properties are familiar. It is often convenient to transform between bare and dressed states
\begin{equation}\label{dress2}
\begin{array}{c}
|\phi_\nu(k)\rangle=\sum_\mu\,c_\nu^\mu(k)|\psi_\mu(k)\rangle \\ \\
|\psi_\mu(k)\rangle=\sum_\nu\,c_\mu^\nu(k)|\phi_\nu(k)\rangle,
\end{array}
\end{equation}
where the coefficients $c_\nu^\mu(k)$ depend on $k$.

\begin{figure}[ht]
\centerline{\includegraphics[width=8cm]{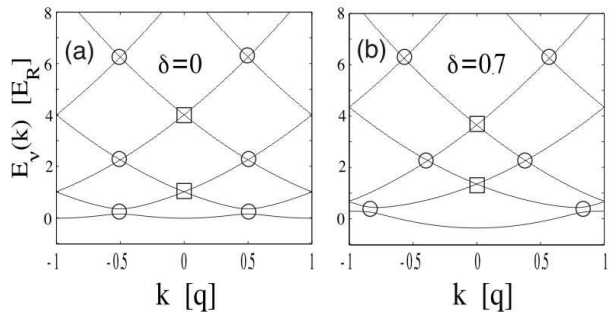}}
\caption[bandstructure]{\label{fig1} The lowest energy bands of the Hamiltonian (\ref{ham}) as function of quasi momentum $k$ for two different detunings $\delta$. The quasi momentum $k$ runs between -1 and 1 in the plots. The Bragg scattering crossings are marked with circles and the Doppleron with square boxes. The coupling is $V_0=0.2$ in both plots.}
\end{figure}

The Hamiltonian (\ref{ham}) has close similarities to the Mathiue equation \cite{math}, and in the $\delta=0$ limit our 'two-level' Hamiltonian is, actually, separable by a unitary transformation  into two uncoupled Mathiue equations. In the opposite limit, $g_0/\delta\rightarrow0$, the Hamiltonian is also diagonal in its internal 'two-level' structure. For the two disconnected equations obtained, the 'potentials' go as $\sim\cos^2(x)$, see \cite{atombloch2}. The 'potentials'  equal $\frac{1}{2}\left[1+\cos(2x)\right]$, where the first term describe the situation when the two photon kicks have opposite momenta, while the second term gives the processes when the absorbed and emitted photons kick the atom in the same direction, either left or right.  In Fig.~1 two examples of the lowest lying bands of the spectrum $E_\nu(k)$ for $-1<k<1$ are shown. In the first plot (a) the detuning is zero, $\delta=0$, while the second plot (b) has a positive detuning $\delta=0.7$. The diagonal matrix elements of the Hamiltonian, in the bare state basis, are $\varepsilon_\mu(k)=(k+\mu)^2\pm\delta/2$, which are, of course, eigenenergies for zero coupling, and therefore called {\it bare eigenenergies}. For a non-zero coupling, the degeneracies are split, forming the gaps in the spectrum. These avoided crossings will be of importance in the remaining of the paper. In the two plots we see two different kinds of crossings (there exist also a third type discussed in section IV), those when both of the bare crossing energies have the same internal state $|\pm\rangle$ and those when they have different internal states. In the first case we say that the crossing point is a Bragg resonance and the second is called a Doppleron resonance \cite{dyn3}. Note that in (b) the most pronounced avoided crossing is between the second and the third band and not between the first and the second band. This is because the lowest bare eigenenergies that build up the two lowest energy bands both belong to states with the same internal state, and these states are only indirectly coupled by the Hamiltonian (\ref{ham}). By controlling the detuning a varity of different spectra can be obtained. For example, the location of the gaps and also  their size can be controlled to some degree. 

The spectrum or the {\it dispersion curves} $E_\nu(k)$, connecting quasi momentum $k$ with energy for a given band, contain information about the dynamics of the system. More precisely, the dispersion curves identify the behaviour of an atomic Gaussian wave packet in the presence of the periodic light field. For us, there are two types of wave packets that are of interest; Gaussian dressed or bare state wave packets. The first one is a Gaussian wave packet built out of dressed states according to
\begin{equation}\label{dressgauss}
|\Phi_{\nu}\rangle=\int_{-1}^{1}dk\,\varphi_\nu(k)|\phi_\nu(k)\rangle,
\end{equation}
where $\varphi_\nu(k)$ is Gaussian centered around some quasi momentum $k_0$ within the first Brillouin zone. Note that we have assumed only a single band to be occupied, thus we do not sum over $\nu$. The second Gaussian state is
\begin{equation}\label{baregauss}
|\Psi\rangle=\int dp\,\chi(p)|p\rangle|-\rangle=\sum_\mu\int_{-1}^{1}dk\,\chi(k+\mu)|\psi_\mu(k)\rangle,
\end{equation}
and $\chi(p)$ is a normalized Gaussian distribution with, for simplicity, a spread within the size of one Brillouin zone. Using Eq.~(\ref{dress2}) for going between the two basis, it follows that $\chi(k)=\varphi_\nu(k-\mu)c_\nu^\mu(k-\mu)$. If $c_\nu^\mu(k)$ is smoothly varying over the spread of the Gaussian this means that a Gaussian distribution in quasi momentum space gives a set  of nearly Gaussian distributions, each shifted by unity in either direction, in real momentum space. The time evolution of a dressed Gaussian wave packet is
\begin{equation}\label{timedressgauss}
|\Phi_\nu(t)\rangle=\int_{-1}^{1}dk\,\varphi_\nu(k)\,\mathrm{e}^{-iE_\nu(k)t}|\phi_\nu(k)\rangle.
\end{equation}
By expanding the dispersion curve around $k_0$ and only including the first three terms (assuming $E_\nu(k)$ to vary little within the spread of $\varphi_\nu(k)$), one may interpret the individual terms of the expansion: The first gives an overall phase-shift, the second and third ones are connected with group velocity and effective mass $m_2$ as
\begin{equation}\label{eff}
\begin{array}{lcccl}
v_g=\left.\displaystyle\frac{\partial E_\nu(k)}{\partial k}\right|_{k=k_0} & & \mathrm{and} & &  
\displaystyle\frac{1}{m_2}=\left.\displaystyle\frac{\partial^2 E_\nu(k)}{\partial k^2}\right|_{k=k_0}.
\end{array}
\end{equation}
In the absent of external forces, $v_g$ determines the velocity of the wave packet and $m_2$ determines how 'fast' the wave packet spreads, see \cite{dyn0} for a detailed discussion. The wave packet thus behaves as freely evolving but with some effective parameters. This is, of course, all well known from the theory of electrons in crystal lattices.

\section{Landau-Zener transitions and Bloch oscillations}\label{sec3}
A very interesting phenomenon of the atomic motion occurs when a linear force is allowed to act on the atom, i.e. the Hamiltonian (\ref{ham}) will include an extra term $\tilde{F}\tilde{x}$, where the coefficient in scaled dimensionless units becomes $F=\tilde{F}/qE_R$. This could, for example, be the force felt by a trapped atom due to gravity. Clearly, by including the extra term, the periodicity is violated and the new eigenstates will have different properties than the former dressed states. The spectrum no longer consists of bands, but becomes purely continuous. It is also commonly known that the spectrum can be described by a discrete set of complex eigenvalues, called Wannier-Stark ladders \cite{wannier}. Contrary to what could be expected, the atomic wave packet will not constantly accelerate, but rather have an oscillatory behaviour. This characteristic has become known as Bloch oscillation originating from \cite{bloch1a,bloch1b}, and it is usually understood either by Wannier-Stark ladders \cite{wannier} or from the typical band structure of fig.~1 \cite{bloch2}. Here we will use the latter of the two approaches to analyze the dynamics. 

When the linear force $F$ is weak enough, an initial Gaussian dressed state will not populate nearby bands. It is the possible to show \cite{bloch1a} that the quasi momentum distribution obeys 
\begin{equation}
\frac{\partial}{\partial t}|\varphi(k,t)|^2=-F\frac{\partial}{\partial k}|\varphi(k,t)|^2,
\end{equation}
with solution
\begin{equation}
|\varphi(k,t)|^2=|\varphi(k-Ft)|^2.
\end{equation}
This is nothing but having an 'adiabatic' evolution. Within this limit, a Gaussian dressed state will move its mean according to $k=k_0-Ft$, and as it exits one Brillouin zone it enters the same zone on the opposite side. All population remains in the same band as the initial state during its evolution, and we see from fig.~1 that, since the group velocity $v_g$ defined in eq.~(\ref{eff}) will flip sign while traversing the quasi momentum, the atomic wave packet will oscillate in $x$-space.  

We know that the validity of adiabatic approximations is related to the 'distance' between adiabatic energy eigen-curves \cite{ad}. Thus, the approximation that no other bands are populated will, most likely, break down near the avoided crossings. In this section we chose the zero detuning situation, $\delta=0$. By neglecting coupling to other bands except close to a crossing, we may treat the problem as a two-level system with a lowest order effective Schr\"odinge equation
\begin{equation}\label{twolevelS}
i\frac{\partial}{\partial t}\!\left[
\begin{array}{c}   
d_{\mu+1} \\ d_\mu
\end{array}\right]\!\!
=\!\!\left[
\begin{array}{cc}
\!(k_0-Ft+1)^2 & \frac{V_0}{2} \\ \frac{V_0}{2} & (k_0-Ft)^2
\end{array}\!\right]\!\!
\left[
\begin{array}{c}   
d_{\mu+1} \\ d_\mu
\end{array}\right]\!,
\end{equation}
where we have assumed a Doppleron resonance crossing. Here $d_\mu$ is the amplitude for the bare state $|\psi_\mu(k_0-Ft)\rangle$. Thus, if we consider bands $\nu$ and $\nu+1$, the various amplitudes are
\begin{equation}\label{amp}
\begin{array}{c}
d_{\mu+1}=\left\{
\begin{array}{lll}
c_\nu^{\mu+1}, & & \mathrm{before\,\,\, crossing}\\
c_{\nu+1}^{\mu+1}, & & \mathrm{after\,\,\, crossing}
\end{array}\right. \\ \\
d_{\mu}=\left\{
\begin{array}{lll}
c_{\nu+1}^{\mu}, & & \mathrm{before\,\,\, crossing}\\
c_{\nu}^{\mu}, & & \mathrm{after\,\,\, crossing}
\end{array}\right.
\end{array}
\end{equation} 
By linearalizing the energy dispersion curves around the first crossing and omitting the constant energy-shifts, we obtain
\begin{equation}\label{LZ}
i\frac{\partial}{\partial t}\left[
\begin{array}{c}   
d_{\mu+1} \\ d_\mu
\end{array}\right]
=\left[
\begin{array}{cc}
F t & \frac{V_0}{2} \\ \frac{V_0}{2} & -F t
\end{array}\right]
\left[
\begin{array}{c}   
d_{\mu+1} \\ d_\mu
\end{array}\right].
\end{equation}
This is the analytically solvable Landau-Zener model \cite{lz1,lz2}, which has been widely used in different areas of physics. The asymptotic solution at $t=\infty$ is
\begin{equation}\label{lzsol}
|d_{\mu}|^2=1-\mathrm{e}^{-\Lambda},
\end{equation}
where the initial condition at $t=-\infty$ is  $|d_{\mu+1}|=1$ and $\Lambda=\pi V_0^2/4F$ is the adiabaticity parameter. The breakdown of not having exact Landau-Zener transitions between neighboring states in the Bloch oscillation scheme has been thoroughly studied in \cite{lzbreak}, and will not be the subject of this paper. The solution ({\ref{lzsol}) is only an approximation, but, nevertheless, it gives a rough idea of the parameter dependence; a small force $F$, the velocity of the momentum wave packet is low, and a strong coupling $V_0$, the band gap is large and the energy curves are far apart, which gives an adiabatic transition between the two bare states $|\psi_\mu(k_0+Ft)\rangle$ and $|\psi_{\mu+1}(k_0+Ft)\rangle$. Thus, according to eq.~(\ref{amp}), no transition takes place between the bands in such a situation. Population that escapes into higher bands due to non adiabatic evolution, will move more freely, since the higher bands are less coupled to one another. Note that in the Landau-Zener solution, the system is integrated between $-\infty$ to $+\infty$ and it is therefore not fully accurate to identify the amplitudes $|d_\mu|^2$ with the ones in the Landau-Zener model. 

\begin{figure}[ht]
\centerline{\includegraphics[width=8cm]{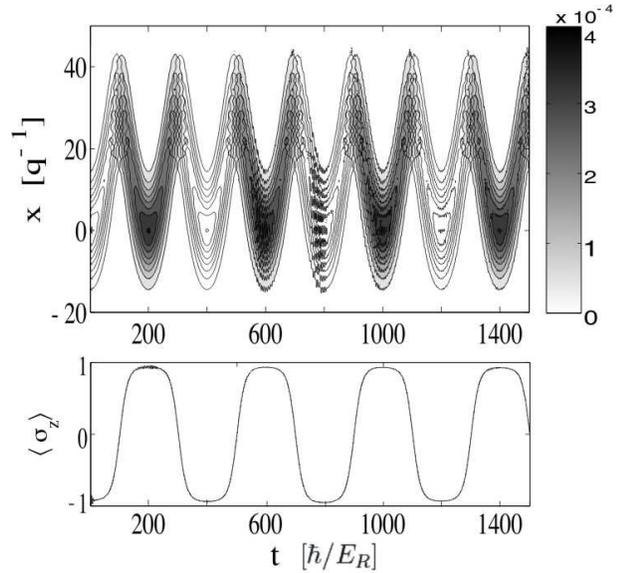}}
\caption[bandstructure]{\label{fig2} Bloch oscillations of an initial Gaussian bare state in the lowest band for zero detuning $\delta=0$. Contours with filled gray scales show the upper atomic wave packet and the non-filled contours the lower wave packet. The scaled coupling is rather weak in this example, $V_0=0.2$, implying that bare and dressed states may have a large overlap. Thus, as the Gaussian quasi-momentum wave packet traverses the lowest band, the atom will Doppleron scatter at each Brillouin zone and therefore the internal state $|\pm\rangle$ is flipped, which is seen in the lower plot showing the atomic inversion $\langle\sigma_z\rangle$. Here the force $F=0.005$.}
\end{figure}

In fig.~2 we show the evolution of an initial Gaussian bare state in the presence of a force $F$. The atom is initially in its lower state with a spatial distribution 
\begin{equation}
\tilde{\chi}(x)=\frac{1}{\sqrt[4]{2\pi\Delta_x^2}}\mathrm{e}^{-\frac{x^2}{4\Delta_x^2}},
\end{equation}
with the width $\Delta_x^2=50$. Thus, it has an average initial quasi momentum $k_0=0$ and in $x$-space it extends over several periods of the standing wave meaning that $\Delta_k^2$ is much smaller than the Brillouin zone. Note that, since the initial state is a bare one, some of the higher bands will be populated already before the first crossing. However, the coupling is weak enough that only a few percentage of the population is in higher dressed states. This is seen in the lower plot in fig.~2 where the atomic inversion $\langle \sigma_z\rangle$ is shown as function of the scaled time. From the atomic inversion we see that the atom 'jumps' between upper and lower states and we also note that the evolution is very adiabatic and not much population leaves the lowest band. As argued above, a larger force $F$ should make the population of the oscillating part damp out. This is seen in fig.~3 where we plot the same as in fig.~2 except that $F$ is three times as large. Note that the Bloch period $T_B=1/F$ is a third as long, and how the atomic inversion damps out much faster than in fig.~2 due to the non-adiabatic transitions.   

\begin{figure}[ht]
\centerline{\includegraphics[width=8cm]{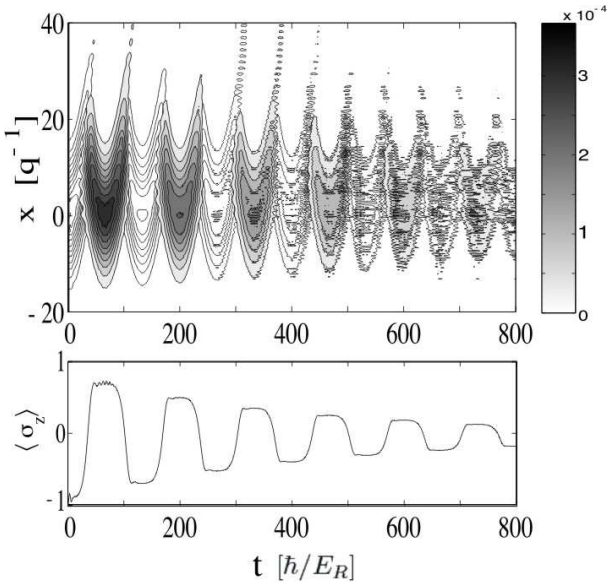}}
\caption[bandstructure]{\label{fig3} This shows the evolution of the same initial state as in fig.~2 except that the force is three times as large, $F=0.015$. The non-adiabatic transitions due to the increased force are seen both in the atomic wave packets and in the atomic inversion, which damps out at each transition.}
\end{figure}
          
\section{Three-level Landau-Zener transitions and Bloch Oscillations}\label{sec4}
The previous section dealt with the dynamics of a wave packet, evolving under the Hamiltonian (\ref{ham}) with the additional 'force' potential $Fx$, in the situation of zero detuning $\delta=0$. Loosely speaking, as the wave packet approached a level crossing, it adiabatically absorbed or emitted a photon, resulting in a flip of internal state and a momentum 'kick'. We said that the atom was Doppleron scattered. The main difference between those Bloch oscillations and the ones most often studied is the internal structure of the system; with no internal structure, the adiabatic transition only results in two or no momentum 'kicks' and no flip, which is that of Bragg scattering.

The existence of Doppleron resonances in this model is just one example of how this model behaves differently from the standard systems lacking internal structure of the particle. By varying the detuning, a third kind of resonance can be achieved, which is seen in fig.~4. In the figure $\delta=1$ or $\delta=-1$, and in both situations there is a level crossing between three bare energy curves. This scattering contains zero, one or two photon exchanges, giving a combination of both Bragg and Doppleron resonances. More generally we have a three level crossing for $\delta=(2j+1)$, $j=...\,,\,-1,\,0,\,1,\,...\,$, where negative detunings are located at $k=0$ and positive ones at $k=\pm1$.
\begin{figure}[ht]
\centerline{\includegraphics[width=8cm]{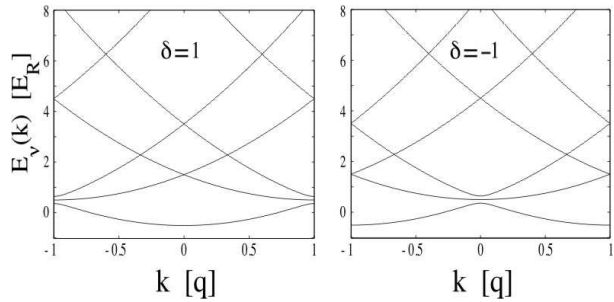}}
\caption[bandstructure]{\label{fig4} The same energy bands as in fig.~1, but for two different detunings. For these detunings, $\delta=1$, -1, there is a new kind of level crossing where three curves intersect. Hence, these crossings are neither purely Bragg nor Doppleron resonances.}
\end{figure}

The evolution of an initial Gaussian bare state with the atom in its lower state when $\delta=1$ is shown in fig.~5. From the atomic inversion, we again note that most of the population is within the lowest band. However, as the wave packet traverses through a crossing, the upper atomic state is being populated. This does not mean that the second energy band is populated, but that close to a crossing the dressed state is a linear combination of bare states containing both upper and lower atomic states. The parameters are the same as in the earlier figs.~2 and 3, except that the force is weaker $F=0.0025$. 

\begin{figure}[ht]
\centerline{\includegraphics[width=8cm]{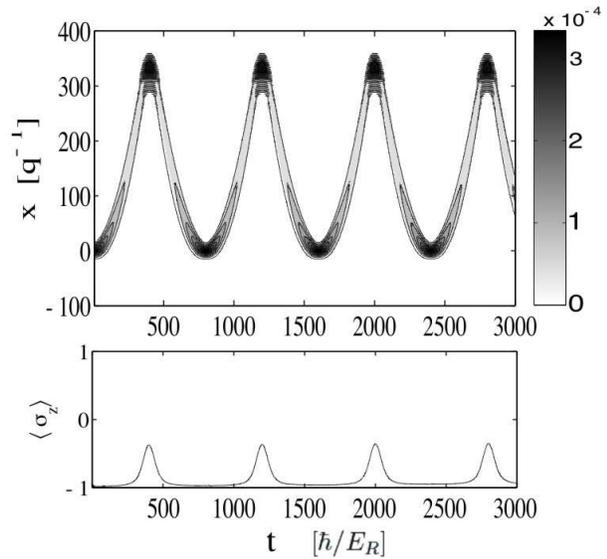}}
\caption[bandstructure]{\label{fig5} Bloch oscillations of an initial Gaussian bare state wave packet  for $\delta=1$ and $k_0=0$. Filled contour curves represent the atom in its lower state, while the non-filled curves (hardly seen in the plot) shows the upper atomic wave packet. From the lower plot, showing the atomic inversion, we note that the upper atomic state is populated almost only close to the curve crossings, see fig.~4. This does, however, not mean that the second band is populated when the quasi momentum wave packet passes a crossing, rather that the ground dressed state is a linear combination of both upper and lower atomic states close to a crossing. Other parameters are, $F=0.0025$, $V_0=0.2$ and $\Delta_x^2=50$.}
\end{figure}

In order to get a deeper insight into the transition we may, similarly to the previous section, treat the Hamiltonian as an effective time dependent three level system by neglecting all other levels and putting $k=k_0-Ft$. Let us shift the crossing to $t=0$ and the energy $E=0$, and then we get the linearalized Hamiltonian
\begin{equation}\label{lz3ham}
H_l=\left[
\begin{array}{ccc}
2Ft & \frac{V_0}{2} & 0 \\ \frac{V_0}{2} & 0 & \frac{V_0}{2} \\ 0 & \frac{V_0}{2} &-2Ft
\end{array}
\right].
\end{equation}
This is a three-level version of the Landau-Zener model and it has been solved analytically in a more generalized form in \cite{lz3}. I may introduce a transition matrix, $\mathcal{T}$, that takes the initial probability amplitudes squared at $t=-\infty$ to their final values at $t=\infty$, provided that only one level is populated originally. In the above model, bare and dressed states become identical at $t=\pm\infty$, except for a flip in subscripts $1\leftrightarrow3$, and the matrix $\mathcal{T}$ has identical elements irrespective if it is given in bare or dressed basis. Thus we have for bare states
\begin{equation}\label{lz3res}
\begin{array}{lll}
\!\left[\!\begin{array}{c}
|d_{\mu-1}^{\infty}|^2 \\  |d_\mu^{\infty}|^2 \\ |d_{\mu+1}^{\infty}|^2\end{array}\!\right]\!\!
 &\!\! =\!\! &\!\! 
\left[\!\!\begin{array}{ccc}
P^2 & 2P(1-P) & (1-P)^2 \\
2P(1-P) & (1-2P)^2 & 2P(1-P) \\
(1-P)^2 & 2P(1-P) & P^2\end{array}\!\!\right]\!\!\times \\ \\ & & 
\left[\begin{array}{c}
|d_{\mu-1}^{-\infty}|^2 \\  |d_\mu^{-\infty}|^2 \\ |d_{\mu+1}^{-\infty}|^2\end{array}\right],
\end{array}
\end{equation}
where only one of the three levels is populated at time $t=-\infty$. The probability is given by
\begin{equation}
P=\mathrm{e}^{-\frac{\Lambda}{2}},
\end{equation}
where the adiabaticity parameter is the same as in the original Landau-Zener model; $\Lambda=\pi V_0^2/4F$. 

\begin{figure}[ht]
\centerline{\includegraphics[width=8cm]{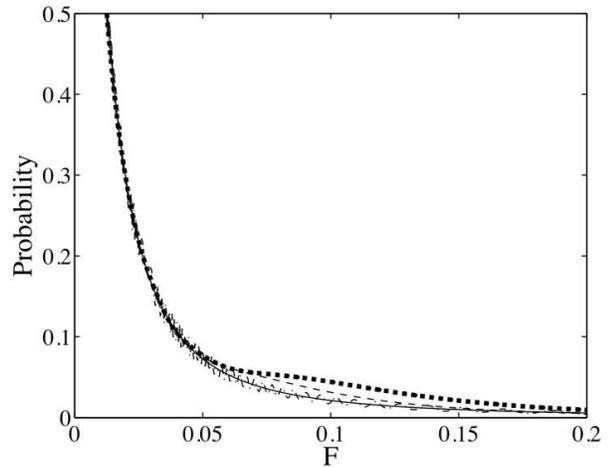}}
\caption[bandstructure]{\label{fig6} This figure shows the probability (\ref{probare}) for the atom to be in its excited state $|+\rangle$ and with momentum $-1<p<0$ as function of force $F$ (dotted line), after an initial Gaussian bare wave packet in the $|+\rangle$ state has been propagated across the crossing at $k=0$ seen in fig.~4. Dashed line displays the probability (\ref{prodress}) of ending up in the lowest band $\nu=1$. The solid curve gives the result of eq.~(\ref{lz3res}) with initial condition $|d_{\mu+1}^{-\infty}|=1$. Finally the dotted-dashed line represents the result from numerical integration of eq.~(\ref{lz3ham}) between $t=-\tau$ and $t=\tau$ corresponding to one half Brillouin zone (from $k_0=-1/2$ to $k_0=1/2$). All four curves approaches zero when the force is increased, leading to a larger non-adiabatic transitions. The other parameters are $V_0=0.2$, $k_0=-0.5$ and $\Delta_x^2=50$. }
\end{figure}

A comment is in order about the result (\ref{lz3res}). In the limit of $t\rightarrow\pm\infty$, the eigenvalues of the Hamiltonian (\ref{lz3ham}) are $-\infty,0,\infty$ and the three states do not couple. However, with the full Hamiltonian (\ref{ham}), in the approximation $k=k_0-Ft$, the crossings are periodic and the adiabatic eigenenergies are never infinitely far apart, so the states will always be coupled. If we start with an initial bare state wave packet in the middle between two crossings, corresponding to a time $t=-\tau$, the overlap with dressed states will be non-zero for several states. At $t=-\tau$, the state is propagated across the crossing till it reaches the next middle point at $t=+\tau$, and there the populations of different bare or dressed states are calculated. In the limit of weak coupling we may expect that only the three lowest bands will be populated after the passage of the crossing, assuming that the initial bare state corresponded to the lowest ground dressed state. These three probabilities are compared with the asymptotic ones $|d_{\mu-1,\mu,\mu+1}^{\infty}|^2$ obtained from eq. (\ref{lz3res}). One may expect that it would be more correct to compare them with the corresponding probabilities obtained from integrating the eq. (\ref{lz3ham}) between $-\tau$ and $+\tau$. The corrections of integrating over a finite time interval depends, of course, on how fast the solutions approache their asymptotic limits, for a thorough discussion see \cite{lzbreak}. In fig.~6, the dotted line gives the probability 
\begin{equation}\label{probare}
P_+(-1<p<0)=\int_{-1}^0dp\,|\langle p,+|\Psi(t=\tau)\rangle|^2,
\end{equation}
where $|\Psi(t=\tau)\rangle$ is the final state propagated across the crossing at $k=0$ shown in the right plot of fig.~4, as a function of the force $F$. The initial state is a Gaussian bare state with internal state $|+\rangle$ and $k_0=-1/2$. Dashed line shows the probability of remaining in the lowest band
\begin{equation}\label{prodress}
P(\nu=1)=\int_{-1}^1dk\,|\langle\phi_{\nu=1}(k)|\Psi(t=\tau)\rangle|^2.
\end{equation}
Finally the solid line and dotted-dashed curve display $|d_{\mu-1}^{\infty}|^2$ obtained with the transition matrix according to eq.~(\ref{lz3res}) and from numerical integration of (\ref{lz3ham}) between $t=\pm\tau$ respectively. It is seen that the probability from numerical integration follows the asymptotic solution very well. The coupling is rather weak in this example, implying that bare and dressed states have a large overlap.       

I do not go into details in understanding the differences between the three curves. In principle a similar analysis as the one performed in \cite{lzbreak} for the original Landau-Zener result could be done. However, the Dykhne-Davis-Pechukas method \cite{ddp1,ddp2} for calculating transition probabilities for two-level systems, which is used in \cite{lzbreak} is, obviously, not justified on our three-level system.

\section{Chirped adiabatic transitions}\label{sec5}
In the previous two sections the atom, while interacting with the standing wave field, was exposed to a constant force. The effect of this force could, in some linear or adiabatic regime, be effectively described by quasi momenta growing linearly in time. From the effective system Hamiltonian we could understand how the dynamics of the atom behaves and how the transitions between states take place. Instead of adding a force $F$ to the system and consequently introducing a time dependence, we could externally vary either the coupling $V_0$ or the detuning $\delta$ during the interaction in order to insert explicit time-dependence. In this section we no longer apply an external force, but instead assume the detuning to be time dependent, and the change in time will be assumed slow, such that we remain in the adiabatic regime throughout the evolution. 

\begin{figure}[ht]
\centerline{\includegraphics[width=8cm]{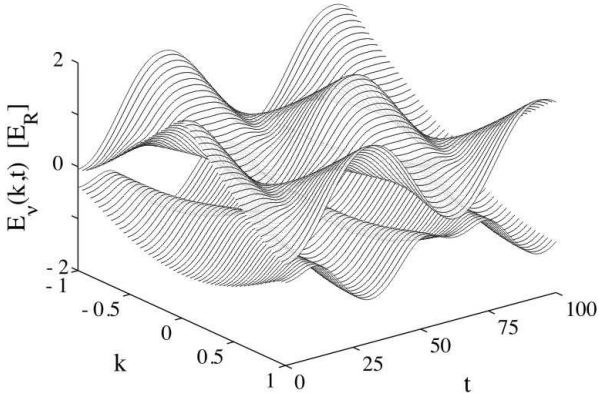}}
\caption[bandstructure]{\label{fig7} The two lowest adiabatic energy bands $E_\nu(k,t)$, obtained by diagonalization of the Hamiltonian with a time dependent detuning $\delta(t)=\delta_0\cos(\omega t)$. It is seen that all crossings are avoided. The parameters are $V_0=0.5$, $\delta_0=2$ and $\omega=0.1$.}
\end{figure}

One interesting choice of a time dependent detuning would be an oscillating one, $\delta(t)=\delta_0\cos(\omega t)$, where $\omega$ sets the time-scale. One could achieve this kind of detuning by, for example, Stark-shift the atom with an external alternating electric field. Usually the Stark-shift is quadratic in the field amplitude, so in order to have the above time-dependence the field variations must be chosen carefully. By making an instantaneous diagonalization of the time-dependent Hamiltonian, we find time dependent adiabatic eigenvalues, $E_\nu(k,t)$. In fig.~7 we plot the time variation of the two lowest bands, for a fairly low detuning amplitude $\delta_0=2$. For an adiabatic process, a Gaussian dressed wave packet will stay within one band and evolve according to the effective parameters (\ref{eff}), which will, of course, change in time.    

Assuming that we can approximate the system to contain just the two lowest dressed states (and $k>0$) we have the effective Hamiltonian
\begin{equation}
H_e=\left[\begin{array}{cc}
(k-1)^2+\frac{\delta(t)}{2} & \frac{V_0}{2} \\
\frac{V_0}{2} & k^2-\frac{\delta(t)}{2}
\end{array}\right],
\end{equation}
and for simplicity we pick $k_0=0.5$, giving the Schr\"odinger equation
\begin{equation}\label{gen2}
i\frac{\partial}{\partial t}\left[\begin{array}{c} \varphi_2 \\ \varphi_1\end{array}\right]=\left[\begin{array}{cc} \frac{\delta(t)}{2} & \frac{V_0}{2} \\ \frac{V_0}{2} & -\frac{\delta(t)}{2}\end{array}\right]\left[\begin{array}{c} \varphi_2 \\ \varphi_1\end{array}\right].
\end{equation}
The eigenvalues of the above Hamiltonian are $\pm\frac{1}{2}\sqrt{\delta^2(t)+V_0^2}$ and thus, when $\delta(t)$ changes sign we have an avoided level crossing. In the adiabatic limit, the population is transfered between the bare states across the crossing. The Schr\"odinger equation (\ref{gen2}) with $\delta(t)=\delta_0\cos(\omega t)$ has been studied in \cite{bmgss}. The adiabatic solution, assuming an infinitely slow change, for the inversion is \cite{bmgnvv}
 \begin{equation}\label{adinv}
\begin{array}{lll}
\langle\sigma_z\rangle & = & 1-\displaystyle{\frac{\delta^2(t)}{\left[V_0^2+\delta_0^2\right]^{1/2}\left[V_0^2+\delta^2(t)\right]^{1/2}}}\\ \\ & & 
-\displaystyle{\frac{V_0^2\cos\left(\phi(t)\right)}{\left[V_0^2+\delta_0^2\right]^{1/2}\left[V_0^2+\delta^2(t)\right]^{1/2}}},
\end{array}
\end{equation}
where
\begin{equation}
\phi(t)=\int_0^t\,dt'\sqrt{V_0^2+\delta^2(t')}
\end{equation}
and with $\delta(t)=\delta_0\cos(\omega t)$, we can express this angle as an elliptical integral 
\begin{equation}
\begin{array}{lccr}
\displaystyle{\phi(t)=V_0\frac{\mathrm{E}(\omega t|m)}{\sqrt{1-m}}}, & & & \displaystyle{\frac{m}{m-1}=\frac{\delta_0^2}{V_0^2}},
\end{array}
\end{equation}
where $\mathrm{E}(\omega t|m)$ is the elliptic integral of the second kind \cite{math}. In the adiabatic limit and first assuming $\cos\left(\phi(t)\right)=1$, the inversion (\ref{adinv}) will oscillate with period $t_{\mathrm{osc}}=\pi/\omega$. The oscillations coming from $\cos\left(\phi(t)\right)$ will be seen as ripples on top of the other, slower oscillations, and the amplitudes of the two kinds of oscillations are determined by $V_0$ and $\delta_0$. 

As already mentioned, in the limit of large detuning, bare and dressed states are identical, and since the change is assumed to be adiabatic, the states will remain dressed also for small detunings. In fig.~8 we show the evolution of an initial bare Gaussian wave packet, when the detuning oscillates as $\delta(t)=\delta_0\cos(\omega t)$. But since the initial detuning is rather large, the state has dressed features also when the detuning is small, which is seen in the oscillations of the wave packet, see \cite{bloch2,lzbreak}. From the lower plot one notes that the oscillations of the inversion die out (compared to the stable adiabatic solution (\ref{adinv})), which is due to the non-adiabatic evolution of the wave packet. However, in terms of the transition rate between the two states, the adiabatic solution and the numerical one agree very well. The amplitude of the detuning is here $\delta=80$, and by using typical experimental parameters (see next section) the unscaled amplitude is of the order of $\sim$MHz. Achieving such strong Stark-shifts should be possible \cite{stark}.

\begin{figure}[ht]
\centerline{\includegraphics[width=8cm]{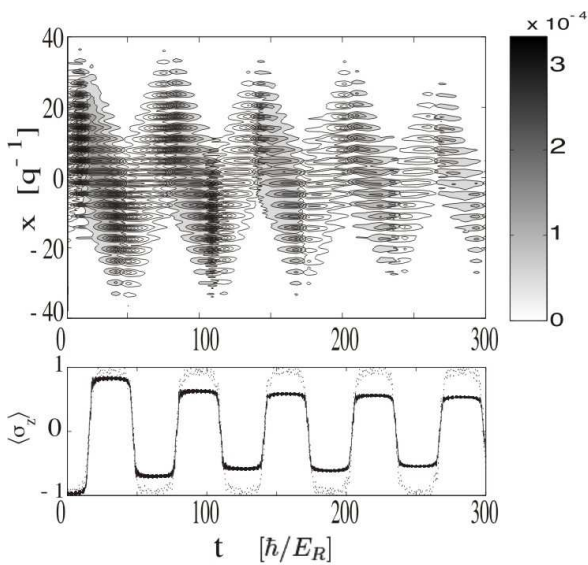}}
\caption[bandstructure]{\label{fig8} The wave packet evolution for the time dependent detuning $\delta(t)=\delta_0\cos(\omega t)$. It is seen that the wave packets moves back and forth as the detuning flips sign. This is due to the adiabatic transition between bare states with different group velocities $v_g$. The oscillations in the wave packet are typical for Gaussian dressed states \cite{bloch2,lzbreak}. In the lower plot, the atomic inversion is shown from the numerical simulation (solid line) and for the adiabatic result (\ref{adinv}) (dotted line). The parameters are $V_0=10$, $\delta_0=80$, $\Delta_x=300$ and $\omega=0.1$.}
\end{figure}

\section{The problem of losses}\label{sec6}

So far losses have been neglected, this is clearly a severe approximation, especially since the dynamics is assumed to be adiabatic. In this section I intend to discuss ways to minimize the losses, and also argue about the effects of losses of the field. 

Throughout the paper I have kept the field quantized, and assumed it to be in a Fock state. The Hamiltonian then has exactly the same form as it would have with a driven field with a large amplitude and the same mode shape. So the analysis is applicable to such a situation as well. One advantage with such fields is that losses are negligible. Thus, losses of the field may be decreased by drive the field with an external classical source. However, in order to have a stable periodic pattern of the field, it should be fixed using some mirrors/cavity. It is understood that since the upper atomic level is populated during the interaction it is most likely that losses will be of great importance. However we may estimate typical time-scales in physical units for the system. In ref.~\cite{atombloch3} the photon recoil energy is $E_R=1.03\times10^{-10}$ eV, which gives a characteristic time $\tilde{t}=\hbar t/E_R\approx1$ ms, where we took $t=200$ from fig.~2. This rather long time may be decreased by using a stronger force $F$, which, however, increase the non-adiabatic contributions. As mentioned, losses of the field is minimized by using a strong driven field while one way to overcome the losses of the excited atomic state is to use a $\Lambda$-type of atom \cite{lambda} as in fig.~9, where the fields are assumed driven. When the detuning $\Delta$ in the figure is large, the upper level can be adiabatically eliminated \cite{adel} to give the interaction Hamiltonian (scaled units)
\begin{equation}
H_{int}=\left[\begin{array}{cc}
\frac{\delta}{2} & V_0\cos(x) \\ V_0\cos(x) & -\frac{\delta}{2}
\end{array}\right],
\end{equation}
where the effective coupling is $V_0=\frac{\Omega_{13}\Omega_{23}}{\Delta}$. Thus, when the elimination of level $|3\rangle$ has been carried out, the dynamics of this model is exactly the same as for the two-level atom used in this paper. 

\begin{figure}[ht]
\centerline{\includegraphics[width=6.5cm]{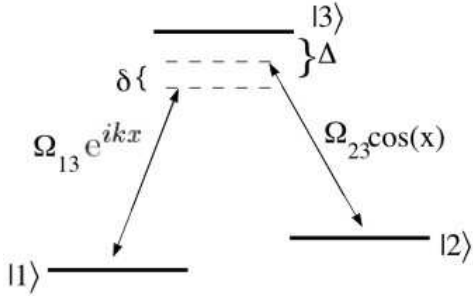}}
\caption[bandstructure]{\label{fig9} $\Lambda$-configuration. The 2-3 transition couples to a standing wave mode, while the 1-3 transition to a traveling wave field.}
\end{figure}

Finally one may use some arguments to get a rough idea about how the losses of the field affects the dynamical quantities. To solve the problem correctly the master equation most be considered. However, we may use some approximations to derive an effective model. By assuming that the field is initially in a coherent state with a large amplitude, it is likely to assume that, in spite of the interaction with the atom, the field remains in some coherent state or in a linear combination of coherent states with the same amplitude. The effect of a zero temperature environment coupled to a coherent state is that the amplitude decrease as $\alpha=\alpha_0\mathrm{e}^{-\kappa t}$, where $\kappa$ is the coupling constant to the environment, see \cite{decay}. As the amplitude of the initial coherent field is large, and hence the photon distribution is sharply peaked around its mean $\bar{n}$, we introduce the effective coupling $V_n(t)=V_0^e\mathrm{e}^{-\kappa t}$ with $V_0^e=V_0\sqrt{\bar{n}}$. Thus, the coupling is assumed to decrease in a similar way as the square root of the mean photon number $\bar{n}$. This is, of course, a rough approximation; the evolution is still coherent, contrary to the one obtained from solving a master equation, and the spread of the photon distribution is not taken into account. However, it should give a measure of typical decoherence times. In fig.~10 I plot the atomic inversion $\langle\sigma_z\rangle$ as a function of time for three different couplings $\kappa^{-1}=250,\,500,\,1000$, where we have assumed an initial Fock state $|1\rangle$ of the field so no $n$ dependence on the coupling. It is seen that, the larger coupling, the faster decay of the inversion;  $\langle\sigma_z\rangle\rightarrow0$. This corresponds to typical decay times 1.5-7 ms. These are rather long for optical cavities ($\sim\mu$s), but not for microwave cavities ($\sim10$ ms). In order to see the effect of the width of the photon distribution I have calculated $\langle\sigma_z\rangle$ for various $n$ (consequently various couplings, chosen as  $V_0\sqrt{n/\bar{n}}\,\mathrm{e}^{-\kappa t}$) and weight them with the distribution of a poisoning (coherent state). This has been done for the examples of fig.~10 with $\bar{n}=50$ and very similar results as in that figure was obtained, and are therefor not shown here. Hence, with initial coherent states with relatively large amplitudes the same kind of evolution as for Fock states is expected.   

\begin{figure}[ht]
\centerline{\includegraphics[width=6.5cm]{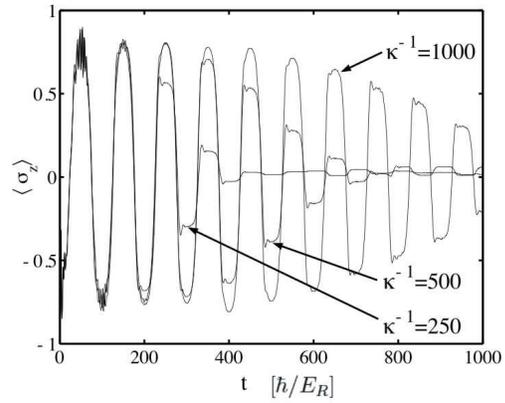}}
\caption[bandstructure]{This shows the results of having a decaying coupling $V_0^e=V_0\mathrm{e}^{-\kappa t}$ for $\kappa^{-1}=250$, 500 and 1000. Here $V_0^e=0.5$, $F=0.02$, $\delta=0$ and $\delta_x^2=50$. The decay of the inversion is seen to increase for larger $\kappa$, as expected. }
\end{figure}

\section{conclusion}\label{concl}
I have analyzed the dynamics of a single two-level atom inside a cavity, interacting with a standing wave mode field. By exposing the atom to a constant force of moderate amplitude it is possible to transform the time independent problem into one where the momentum grows linearly with time. This transformation is only valid for weak forces, suggesting that we can apply the adiabatic theorem, in other words, time can be considered as a parameter and the Hamiltonian can be instantaneously diagonalized. The system, initially in an eigenstate of the Hamiltonian in the absence of the force, will remain in the same energy band even when the force is affecting the dynamics; $|\phi_\nu(k=k_0-Ft)\rangle$. As the quasi momentum is changed, the system traverses so called level crossings where the bare energies cross. It is at these crossings that the adiabaticity constrains are most likely to be violated, meaning that other states/bands may be populated. If the population still remains in a single band even through the passage of a level crossings, the atomic motion will be oscillating, contrary to standard behaviour were the particle accelerates to higher and higher velocities due to the force. This is usually called Bloch oscillations and is a consequence of the energy gaps/crossings in the otherwice continuous spectrum. 

The Bloch oscillations of the atom are investigated for various values of the atom-field detuning. Bloch oscillations of atoms interacting with standing wave fields have long been known and also seen experimentally. But it has almost always been assumed that the detuning is large so that one internal electronic state of the atom can be adiabatically eliminated, and the internal structure is overlooked. Here I have studied the situations when both atomic levels are included, and the differences in the dynamics due to the two-level structure. In the situation when just one level needs to be taken into account, the atom undergoes a two-photon process as it traverses a level crossing adiabatically, it is Bragg scattered. With small detunings, one-photon processes are also possible, making the atom flip its electronic state. In the zero detuning situation, we have seen that for a weak coupling and a small force the atom oscillates both in its motion but also between the two electronic states. We also note how non-adiabatic evolution started to dominate for larger values of the force. The Landau-Zener model was introduced in order to reach a deeper understanding of  the parameter dependences.

For particular detunings a new kind of level crossing is present. In these situations, three bare energy curves cross, and the atom may undergo zero, one or two photon exchanges with the field. The results from numerical wave packet propagations across such a level crossing has been compared with analytic solutions of the three-level Landau-Zener model. The agreement is very good even though there are numbers of approximations made in order to derive the effective three-level Landau-Zener Hamiltonian. 

In the section \ref{sec5}, we omitted the external force, and instead considered the atom-field detuning to be time-dependent $\delta=\delta(t)$. If the detuning changes sign, the bare energy curves, or in this adiabatic case surfaces, cross. By letting the initial Gaussian bare wave packet be centered around some momentum $k_0\neq0$ it has been possible to obtain a similar oscillatory action as in the case with a constant force. The results are compared with the analytically obtained ones in the adiabatic limit.  

The effect of field losses and how to overcome such losses and also atomic losses were discussed in the last section \ref{sec6}. By using an $\Lambda$-type of atom and drive one transition by the cavity field and one by an external classical field an effective two-level atom with reduced losses can be derived by adiabatic elimination of the excited state. The field losses can be decreased by driving it with an external source. I display some numerical results when the field losses are included, and it is clear that the system is very sensitive to losses of realistic sizes. However, using the effective $\Lambda$ model with driven fields, it does not seem too far out of reach to experimentally verify the theoretical predictions in this paper.


\section*{Acknowledgements}
The author wishes to thank Prof. Stig Stenholm and Dr. Janne Salo for intersting and inspiring discussions.


\end{document}